\documentclass[oneside,reqno]{amsart}
\usepackage{amsmath,amsthm,amssymb,amsfonts,mathrsfs,enumerate,graphicx}

\def \reals{{\mathbb R}}

\def \sphere{{\mathbb S}}
\def \supp{\mathop{\rm supp}\nolimits}          

\newcommand{\dummy}{\protect \rule{0.0em}{1.0ex}}
\newcommand{\Dummy}{\protect \rule{0.0em}{1.6ex}}
\newcommand{\preone}{\,\Dummy^{(1)}}
\newcommand{\pretwo}{\,\Dummy^{(2)}}

\def \smallfrac#1#2{{\textstyle \frac{#1}{#2}}}

\newcommand{\g}{{\bf g}}
\newcommand{\bg}{\bar{\g}\dummy}
\newcommand{\tg}{\widetilde{\g} \dummy}

\def \edge{\mathop{\rm edge}\nolimits}
\def \Ricc{\mathop{\rm Ricc}\nolimits}

\newcommand{\barp}{\bar{p}\dummy}
\newcommand{\barq}{\bar{q}\dummy}
\newcommand{\barr}{\bar{r}\dummy}

\newcommand{\barx}{\bar{x}\dummy}
\newcommand{\bary}{\bar{y}\dummy}

\newcommand{\tildep}{\widetilde{p}\dummy}
\newcommand{\tildeq}{\widetilde{q}\dummy}

\newcommand{\tildez}{\widetilde{z}\dummy}

\newcommand{\bD}{\bar{D}\Dummy}

\newcommand{\bI}{\bar{I}\Dummy}

\newcommand{\bM}{\bar{M}\Dummy}

\newcommand{\tM}{\makebox[1em]{\hspace{1pt}$\widetilde{\hspace{0.5pt}%
\makebox[0.9em][r]{$M$}}$}\Dummy}

\newcommand{\dI}{\dot{I}\Dummy}
\newcommand{\dJ}{\dot{J}\Dummy}

\newcommand{\scrC}{\mathscr{C}}

\newcommand{\scrH}{\mathscr{H}}
\newcommand{\scrI}{\mathscr{I}}

\newcommand{\scrK}{\mathscr{K}}

\newcommand{\scrN}{\mathscr{N}}
\newcommand{\scrO}{\mathscr{O}}

\newcommand{\scrQ}{\mathscr{Q}}

\newcommand{\scrS}{\mathscr{S}}
\newcommand{\scrT}{\mathscr{T}}
\newcommand{\scrU}{\mathscr{U}}
\newcommand{\scrV}{\mathscr{V}}

\newcommand{\tscrA}{\makebox[1em]{\hspace{1pt}$\widetilde{\hspace{-3.0pt}%
\makebox[0.9em][r]{$\mathscr{A}$}}$\hspace{-2pt}}\Dummy}

\newcommand{\tscrI}{\makebox[1em]{\hspace{1pt}$\widetilde{\hspace{-3.0pt}%
\makebox[0.9em][r]{$\mathscr{I}$}}$\hspace{-2pt}}\Dummy}

\newcommand{\tscrN}{\makebox[1em]{\hspace{1pt}$\widetilde{\hspace{-3.0pt}%
\makebox[0.9em][r]{$\mathscr{N}$}}$\hspace{-2pt}}\Dummy}

\newcommand{\tscrQ}{\makebox[1em]{\hspace{1pt}$\widetilde{\hspace{-3.0pt}%
\makebox[0.9em][r]{$\mathscr{Q}$}}$\hspace{-2pt}}\Dummy}

\newcommand{\tscrX}{\makebox[1em]{\hspace{1pt}$\widetilde{\hspace{-3.0pt}%
\makebox[0.9em][r]{$\mathscr{X}$}}$\hspace{-2pt}}\Dummy}

\newcommand{\bgamma}{\bar{\gamma}\dummy}

\newcommand{\blambda}{\bar{\lambda}\Dummy}

\newcommand{\talpha}{\widetilde{\alpha}\dummy}

\newcommand{\tgamma}{\widetilde{\gamma}\dummy}

\newcommand{\tlambda}{\widetilde{\lambda}\Dummy}
\newcommand{\tmu}{\widetilde{\mu}\dummy}
\newcommand{\tnu}{\widetilde{\nu}\dummy}

\newcommand{\dgamma}{\dot{\gamma}\dummy}

\newcommand{\tSigma}{\widetilde{\Sigma}\Dummy}

\protect
\protect\newtheorem*{corollary*}{Corollary}
\protect
\protect\newtheorem*{sublemma*}{Sub-lemma}
\protect\newtheorem{theorem}{Theorem{}\protect\setcounter{corollary}{0}%
\setcounter{sublemma}{0}}[section]
\protect\newtheorem*{theorem*}{Theorem{}}
\protect\newtheorem{lemma}[theorem]{\setcounter{corollary}{0}Lemma}
\protect\newtheorem*{lemma*}{\setcounter{corollary}{0}Lemma}
\protect\newtheorem{proposition}[theorem]{\setcounter{corollary}{0}Proposition}
\protect\newtheorem*{proposition*}{\setcounter{corollary}{0}Proposition}
\protect\newtheorem{definition}[theorem]{Definition}
\protect\newtheorem*{definition*}{Definition}
{\protect\theoremstyle{remark}}
{\protect\theoremstyle{remark}\newtheorem*{scholium*}{Scholium}}
{\protect}
{\protect\newtheorem{conjecture*}[theorem]{\bf Conjecture}}
\protect\newtheorem*{remark*}{Remark}

{\theoremstyle{remark}\newtheorem{provisionaldefinition}[theorem]%
{Provisional Definition}}
\renewenvironment{proof}[1][]{\noindent{\em Proof#1. }}%
{\mbox{}\hspace{\fill}\qedsymbol \\[-0.5ex] \mbox{}}
{\mbox{}\hspace{\fill}\qedsymbol \\[-0.5ex] \mbox{}}
\newcommand{\saveqedsymbol}{\qedsymbol}
\newcommand{\subqedsymbol}{\raisebox{0.5ex}{$\bigtriangledown$}}
{\begin{proof}[#1]}%
{\renewcommand{\qedsymbol}{\subqedsymbol}\end{proof}%
\renewcommand{\qedsymbol}{\saveqedsymbol}}

\newcommand{\noproof}{\hspace{\fill}\qedsymbol \vspace{1ex}}
\theoremstyle{remark}
\theoremstyle{remark}
\theoremstyle{remark}
\begin{document}
\pagenumbering{arabic}
\title[Black holes and closed trapped surfaces]{Black holes and
closed trapped surfaces:\\ a revision of a classic theorem}
\author[C.-M. Claudel]{Clarissa-Marie Claudel}
\address{Mathematical Institute, University of Ume{\aa}, S-901 87 Ume{\aa},
Sweden.}
\begin{abstract}
It is standard assertion in relativity that, subject to an energy
condition and the cosmic censorship hypothesis, closed trapped
surfaces are not visible from $\scrI^+$. A proof given by Hawking
\& Ellis in {\em The Large Scale Structure of Space-Time} is
flawed since it is formulated in terms of an inadequate
definition of a weakly asymptotically simple and empty
space-time.  A new proof is given based on a more restrictive
definition of a weakly asymptotically simple and empty space-time.
\end{abstract}
\maketitle
\section{Introduction}
Weakly asymptotically simple and empty (WASE) space-times provide a
setting for the analysis of isolated gravitational objects in
classical general relativity.  There is sufficient freedom and
generality to allow for the presence of singularities and black
holes, but there is also sufficient structure, in the form of a well-behaved
asymptotic region, to permit the proof of significant results.
Examples of WASE space-times include the Schwarzschild, Kerr and
Reissner-Nordstr{\"o}m space-times.

Despite the evident physical importance of WASE space-times, it is
often the case that scant attention is paid to their precise
definition. Indeed, as will be seen, the definition proposed and
used by Hawking \& Ellis \cite{HE73a}, henceforth H{\&}E, is
inadequate for their purposes.  A more restrictive definition was
subsequently proposed in \cite{New84a} as a foundation for a class
of censorship theorems (see also \cite{CF82,New84c,New84d,New84e}).
One purpose of the present paper is to
provide basic results for this definition further to those given
in \cite{New84a}.

As has been well-known since \cite{Pen65a}, the concept of a
closed trapped surface is central to the understanding of black
holes. In the context of WASE space-times it is a standard
assertion, stated as Proposition 9.2.1 in H{\&}E, that subject to
a form of weak cosmic censorship and a suitable energy condition,
it is not possible for closed trapped surfaces to be seen from
future null infinity.  In itself this is of no direct physical
significance because there need be nothing remarkable about the
geometry at a space-time point which lies on a closed trapped
surface. The assertion does however imply, again subject to weak
cosmic censorship and an energy condition, that the presence of a
closed trapped surface in a WASE space-time implies the presence
of a black hole. The assertion also leads to an elementary
censorship theorem according to which, in a WASE satisfying the
same energy condition, if every future singularity is sufficiently
strong as to be preceded by a closed trapped surface, then weak
cosmic censorship must hold.  The second purpose of this paper is
to obtain a rigorous proof of a modified statement of H{\&}E Proposition
9.2.1 on the basis of the definition of a WASE space-time given in
\cite{New84a}.

\section{Preliminary concepts}\label{prelimsec}
The following two basic definitions are taken from \cite{New84a}.

\begin{definition}\label{asymptotedef}
An asymptote of a space-time $(M,\g )$ is a quadruple
$(\bM,\bg,\Omega , \psi )$, where $(\bM,\bg)$ is
a space-time-with-boundary, $\Omega$ is a $C^{\infty}$ real-valued
function on $\bM$, and $\psi : M \rightarrow \bM$ is a
$C^{\infty}$ embedding such that
\begin{enumerate}[(i)]
\item $\psi (M) = \bM \setminus \partial \bM$;
\item $(\psi^*\Omega )^2\g = \psi^*\bg$;
\item one has $\Omega (p) =0$ and ${\bf d}\Omega (p) \neq 0$ for all
$p \in \partial \bM$.
\end{enumerate}
\end{definition}

\begin{definition}\label{ASEdef}
An asymptote $(\tM,\tg,\Omega' , \psi' )$ of a space-time
$(M',\g' )$ is asymptotically simple and empty (ASE) if
\begin{enumerate}[(i)]
\item $\tM \setminus \supp (\psi'_*\Ricc (\g' ))\supset
\partial \tM$;
\item $(\tM,\tg )$ is strongly causal;
\item every inextendible null geodesic $\gamma'$ of $(M',\g' )$
is such that $\psi' \circ \gamma'$ has two endpoints in $\tM$, both
of which lie in $\partial \tM$.
\end{enumerate}
A space-time is asymptotically simple and empty (ASE) if it admits an
asymptotically simple and empty asymptote.
\end{definition}

Standard arguments give that any ASE space-time $(M,\g )$ is
globally hyperbolic, and that for any ASE asymptote
$(\tM,\tg ,\Omega', \psi')$ of $(M',\g' )$, the boundary
$\partial \tM$ of $\tM$ is the union of two disjoint connected null
hypersurfaces $\tscrI^+ := I^+(\tM,\tg;\tM)\cap \partial \tM$ and
$\tscrI^- := I^-(\tM,\tg;\tM)\cap \partial \tM$.
By means of condition (ii) of Definition \ref{ASEdef} one can show
\cite{New89b} that $\tscrI^+$ and $\tscrI^-$ are diffeomorphic to
${\mathbb S}^2\times {\mathbb R}$.

In the terminology of \cite{New89b}, a slice of $\tscrI^+$ is a
non-empty locally acausal compact connected topological
2-submanifold of $\tscrI^+$. Theorem 5.1 of \cite{New89b}
gives that every slice of $\tscrI^+$ is homeomorphic to $\sphere^2$.
Since, by assumption here, strong causality holds at every point of $\tscrI^+$
in $(\tM,\tg)$, Proposition 7.1 of \cite{New89b} gives that every null
geodesic generator of $\tscrI^+$ cuts every slice of $\tscrI^+$,
and Theorem 7.4 of \cite{New89b} gives that $\tscrI^+$ is acausal
in $(\bM,\bg)$.  Any given slice of $\tscrI^+$ may be mapped
along the generators of $\tscrI^+$ to yield a foliation of
$\tscrI^+$ by slices of $\tscrI^+$.
Similar assertions apply to $\tscrI^-$.

Note that $(\tM,\tg)$ need not be causally simple. For example,
let $(M',\g')$ be Minkowski space. Let $\tmu:\reals \rightarrow
\tscrI^-$ be a future-directed null geodesic generator of
$\tscrI^-$ and let $\tnu :\reals \rightarrow \tscrI^+$ be the
antipodal future-directed null geodesic generator of $\tscrI^+$.
Let $\tildep \in|\tmu|$ and let $\tildeq :=
\dJ^+(\tildep,\tg;\tM)\cap |\tnu|\in \tscrI^+$. One then has
$J^+(\tildep,\tg;\tM)\cap \tscrI^+= \tscrI^+\setminus
|\tlambda_{\tildeq}|$ for $\tlambda_{\tildeq} := \tnu |(-\infty
,a)$, for $a:=\tnu^{-1}(\tildeq)\in \reals$. Since
$\tscrI^+\setminus |\tlambda_{\tildeq}|$ is not relatively closed
in $\tscrI^+$, the set $J^+(\tildep,\tg;\tM)$ cannot be closed in
$(\tM,\tg)$.

Condition (ii) of Definition \ref{ASEdef} may be decomposed into
two parts, first that $(M',\g')$ is strongly causal and second
that $(\tM,\tg)$ is strongly causal at every point of
$\tscrI^+$.  Since the physical interpretation of the latter is
unclear one is led, as in \cite{New89b}, to the more general
concept of a ``simple'' space-time for which only the chronology
condition is imposed on $(M',\g')$, with no additional causality
conditions imposed on $\tscrI^+$ or $\tscrI^-$ in $(\tM,\tg)$.
Simple space-times are globally hyperbolic with Cauchy surfaces
which, subject to the truth of the Poincar{\'e} conjecture, are
diffeomorphic to $\reals^3$. But the topological and causal
structure of $\tscrI^+$ and $\tscrI^-$ may exhibit new
complications and subtleties.  Despite the possible interest of
this additional generality, $(\tM,\tg)$ will for present purposes
be assumed to be strongly causal as expressed in condition (ii)
if Definition \ref{ASEdef}.

In H{\&}E, a weakly asymptotically simple and empty (WASE)
space-time is introduced with a definition which, when
re-expressed in terms of asymptotes, assumes the following form.

\begin{provisionaldefinition}[c.f.\ H{\&}E p.225]\label{HEWASE}
A space-time $(M,\g)$ is weakly asymptotically simple and empty
(WASE) if there exists an asymptote $(\bM,\bg,\Omega,\psi )$ of
$(M,\g )$, a space-time $(M',\g')$, an asymptote
$(\tM,\tg,\Omega ', \psi ')$ of $(M',\g')$ and open
sets $\scrU$ and $\scrU'$ of $M$ and $M'$ respectively such that
\begin{enumerate}[(i)]
\item $(\tM,\tg,\Omega',\psi' )$ is ASE;
\item $\psi(\scrU)\cup
\partial \bM$ is an open neighbourhood of $\partial \bM$ in $\bM$;
\item $\psi'(\scrU')\cup \partial \tM$ is
an open neighbourhood of $\partial \tM$ in $\tM$;
\item $(\scrU ,\g |\scrU )$ and $(\scrU',\g'|\scrU')$ are globally
isometric.
\end{enumerate}
\end{provisionaldefinition}

This definition of a weakly asymptotically simple and empty
space-time is not as restrictive as its authors seem to have
intended.  In particular, it allows the future and past null
infinities $\scrI^+:= I^+(\bM,\bg;\bM)\cap \partial \bM$ and
$\scrI^-:= I^-(\bM,\bg;\bM)\cap \partial \bM$ of $(\bM,\bg)$ to be
separated from one another and not join up at a spatial infinity
(see Figure 1). (To add an assumption that $\scrU$ is connected
would not help.) The deficiency undermines several of their
results, and in particular their proposed proof (Proposition
9.2.1) that closed trapped surfaces are necessarily confined to
black holes.
\begin{figure}[h]
\begin{center}
\includegraphics[width=4.0in]{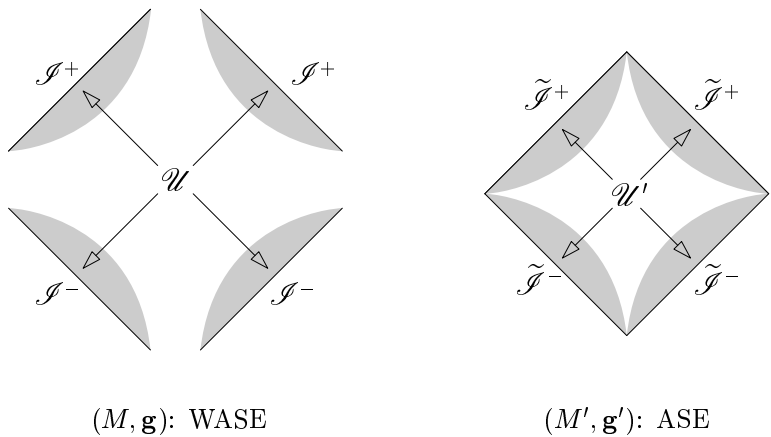}
\end{center}
\caption{The H{\&}E definition of a WASE space-time has the
unwanted feature that the future and past null infinities
$\scrI^+$ and $\scrI^-$ of $(M,\g )$ need not join up at a
spatial infinity.}
\end{figure}

In order to overcome the problems in Definition \ref{HEWASE}, one
may adopt the following, more restrictive definition of a WASE
space-time proposed in \cite{New84a}.

\begin{definition}\label{NWASE}
An asymptote $(\bM,\bg,\Omega,\psi )$ of a space-time $(M,\g )$ is
weakly asymptotically simple and empty (WASE) if there exists an
open set $\scrU$ of $M$, an extension $(M',\g')$ of $(\scrU,\g
|\scrU )$, an asymptote $(\tM,\tg,\Omega',\psi')$ of $(M',\g')$
and a topological embedding $\xi : \overline{\psi (\scrU )\cup
\partial \bM}\rightarrow \tM$ such that
\begin{enumerate}[(i)]
\item $(\tM,\tg,\Omega',\psi')$ is ASE;
\item for every $p' \in M'$, the set $M'\setminus (\scrU \cup
I(p',\g';M'))$ is compact;
\item one has $\xi (\partial \bM )=\partial \tM$ and $\xi \circ \psi
|\scrU =\psi'|\scrU$;
\item for all $q \in \scrU$ and all future-pointing timelike
vectors ${\bf v}\in T_qM$ of $(M,\g )$, the vectors $\psi_*{\bf
v}$, ${\bf v}$ and $\psi'_*{\bf v}$ are future-pointing in
$(\bM,\bg )$, $(M',\g' )$ and $(\tM,\tg )$ respectively.
\end{enumerate}
A space-time is weakly asymptotically simple and empty (WASE) if it
admits a weakly asymptotically simple and empty asymptote.
\end{definition}

\begin{remark*}
For any Cauchy surface $\scrS'$ of $(M',\g')$ one has
$\scrS'\setminus \scrU \subset  M'\setminus (\scrU \cup
I(p',\g';M'))$ for any $p'\in \scrS'$ and hence that
$\scrS'\setminus \scrU$ is compact.
\end{remark*}

The future and past null infinities of a WASE
asymptote $(\bM,\bg,\Omega,\psi )$ are defined by $\scrI^+:= I^+(\bM,\bg
;\bM)\cap \partial \bM$ and $\scrI^-:= I^-(\bM,\bg ;\bM)\cap \partial \bM$
respectively.  It is clear that $\partial \bM$ is the disjoint
union of $\scrI^+$ and $\scrI^-$.

The mappings involved in Definition \ref{NWASE} are shown in
Figure 2.  In Figure 3 one can see how the pathologies inherent in
Definition \ref{HEWASE} are eliminated by condition (ii) of
Definition \ref{NWASE}.  This condition may be regarded as a
way to require that $\scrU$ is a neighbourhood of spatial
infinity without reference to the geometrical structure of spatial
infinity.

\begin{figure}[h]
\begin{center}
\includegraphics[width=3.0in]{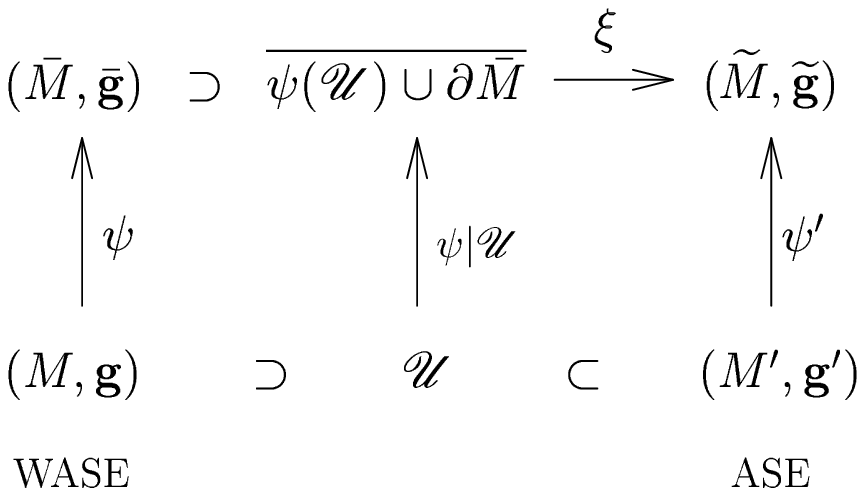}
\end{center}
\caption{The mappings in the Definition \ref{NWASE} of a WASE
space-time.  The left and right squares commute in the category of
topological spaces and continuous mappings.}
\end{figure}

\begin{figure}[h]
\begin{center}
\includegraphics[width=4.5in]{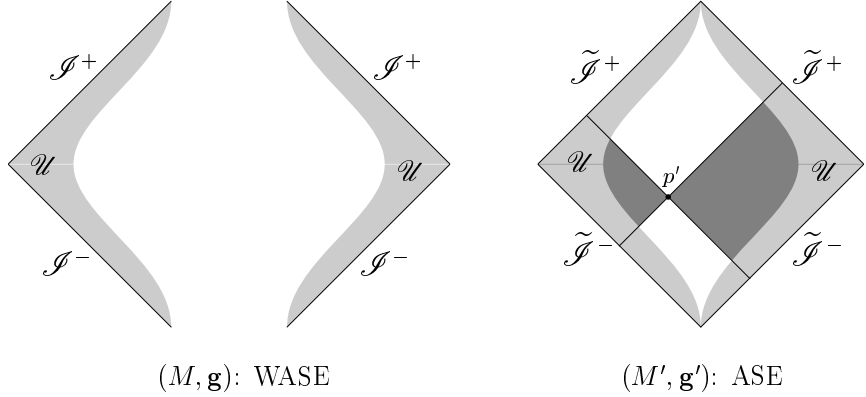}
\end{center}
\caption{The Definition \ref{NWASE} of a WASE space-time requires
that $\scrU$ is such that the darkly shaded region in the right
hand figure is compact for all $p' \in M'$.  This has the desired
consequence that $\scrI^+$ and $\scrI^-$ in the left hand figure
must meet at a spatial infinity.}
\end{figure}

The following two lemmas are basic in the analysis of WASE
space-times.  Their proofs are given in \cite{New84a}.

\begin{lemma}\label{basiclemma1}
Within the context of Definition \ref{NWASE} one has
\begin{enumerate}[(i)]
\item $\psi' (\scrU )\cup \partial \tM$ is open in $\tM$
\item $\psi (\scrU )\cup \partial \bM$ is open in $\bM$.  \noproof
\end{enumerate}
\end{lemma}

\begin{lemma}\label{basiclemma2}
Within the context of Definition \ref{NWASE} one has
\begin{enumerate}[(i)]
\item a subset of $\scrU$ is open in $M$ iff it is open in
$M'$;
\item a subset of $\scrU$ is compact in $M$ iff it is compact in
$M'$;
\item a subset of $\scrU$ is closed in $M$ if (but not only if)
it is closed in $M'$. \noproof
\end{enumerate}
\end{lemma}

Since the mappings $\psi :M \rightarrow \bM$ and $\psi' : M'\rightarrow
\tM$ in Definition \ref{asymptotedef} are both diffeomorphisms onto
their images, it is clear from Figure 2 that the mapping $\xi
:\overline{\psi (\scrU )\cup \bM}\rightarrow \tM$ in Definition
\ref{NWASE} is such that $\xi |\psi (\scrU )$ is a diffeomorphism
onto $\psi'(\scrU )$.  Moreover $\xi|\psi(\scrU)$ is a conformal isometry
onto $\psi'(\scrU)$ in the sense of $(\xi |\psi (\scrU))^*\tg =
(\Omega /\Omega')^2\bg | \psi (\scrU )$.
Note however that $\xi$ need not be differentiable at points
of $\partial \bM$.

\begin{proposition}\label{generators}
One has
\begin{enumerate}[(i)]
\item $\xi (\scrI^+) = \tscrI^+$ and
$\xi (\scrI^-) = \tscrI^-$;
\item $\xi$ maps the null geodesic generators of $\scrI^+$ and $\scrI^-$
onto null geodesic generators of $\tscrI^+$ and $\tscrI^-$
respectively.
\end{enumerate}
\end{proposition}

\begin{proof}
Let $p \in \scrI^+$.  Lemma \ref{basiclemma1} gives that
$\psi(\scrU)\cup \partial \bM$ is an open neighbourhood of $p$ in
$\bM$. Hence, by the definition of $\scrI^+$, there exists a
smooth timelike curve $\alpha :(0,1)\rightarrow \psi (\scrU )$ of
$(\bM ,\bg )$ having a future endpoint at $p$ in $(\bM ,\bg )$.
One has $\xi (p)\in \partial \tM$ by condition (iii) of Definition
\ref{NWASE}. Since $\xi \circ \alpha$ is a smooth timelike curve
of $(\tM,\tg )$ with a future endpoint at $\xi (p)$ one therefore
has $\xi (p)\in \tscrI^+$ by the definition of $\tscrI^+$.  There
follows $\xi (\scrI^+)\subset \tscrI^+$. A similar argument gives
$\xi^{-1}(\tscrI^+) \subset \scrI^+$ which implies $\tscrI^+
\subset \xi (\scrI^+)$.  Hence one has $\xi (\scrI^+) = \tscrI^+$
and similarly $\xi (\scrI^-) = \tscrI^-$. This establishes (i).

Let $\gamma :\reals \supset I \rightarrow \scrI^+$ be a
future-directed null geodesic generator of $\scrI^+$ and let
$\tgamma:= \xi \circ \gamma$.  Let $[a,b]\subset I$ for $a<b$ and
let $\mu:=\gamma|[a,b)$. One has $|\mu|\subset \scrI^+$. By Lemma
\ref{basiclemma1} and the definition of $\scrI^+$ there exists a
smooth future-directed timelike curve $\mu_0:[a,b)\rightarrow
\psi(\scrU)$ of $(\bM ,\bg )$ with a future endpoint at $\gamma
(b)$. Indeed there exists a sequence of smooth future-directed
timelike curves $\mu_i :[a,b)\rightarrow \psi(\scrU)$ of $(\bM
,\bg )$ converging pointwise to $\mu$ in $\bM$, each with a
future endpoint at $\gamma (b)$. Since the smooth timelike curves
$\tmu_i:=\xi \circ \mu_i:[a,b) \rightarrow \psi'(\scrU)$ of
$(\tM,\tg)$ converge pointwise to $\tmu :=\xi \circ \mu$ in $\tM$
one has that $\tmu$ is a causal curve of $(\tM,\tg)$ with a
future endpoint at $\tgamma (b)$. Since part (i) gives
$|\tmu|=\xi (|\mu|)\subset \tscrI^+$ and since $\tscrI^+$ is a
null hypersurface of $(\tM,\tg)$ it follows that $\tmu=\xi \circ
\mu$ is a null geodesic generating segment of $\tscrI^+$. Since
$a$ and $b$ were arbitrary in $\mu:=\gamma|[a,b)$ it follows that
$\tgamma =\xi \circ \gamma$ is a null geodesic generating segment
of $\tscrI^+$. Clearly $\tgamma$ cannot have a future endpoint $q
\in \tscrI^+$ otherwise $\gamma$ would have a future endpoint at
$\xi^{-1}(q)\in \scrI^+$ and so would not be a generator of
$\scrI^+$, contrary to hypothesis. Similarly $\tgamma$ cannot
have a past endpoint in $\tscrI^+$. Hence $\tgamma$ is a null
geodesic generator of $\tscrI^+$. The corresponding result for
$\scrI^-$ is similar. This establishes (ii).
\end{proof}

Proposition \ref{generators} shows that the structure of $\scrI^+$
and $\scrI^-$ for a WASE asymptote $(\bM,\bg,\Omega,\psi)$ is
directly analogous to the structure of $\tscrI^+$ and $\tscrI^-$
for an ASE asymptote $(\tM,\tg,\Omega',\psi)$.  In particular one
may, following \cite{New89b}, define a slice of $\scrI^+$
(respectively a slice of $\scrI^-$) as a non-empty locally
acausal compact connected topological 2-submanifold of $\scrI^+$
(respectively $\scrI^-$). Then $\scrI^+$ and $\scrI^-$ are
acausal in $(\bM,\bg)$, every null geodesic generator of
$\scrI^+$ cuts every slice of $\scrI^+$ and every null geodesic
generator of $\scrI^-$ cuts every slice of $\scrI^-$.  Slices of
$\scrI^+$ and $\scrI^-$ are mapped by $\xi$ to slices of
$\tscrI^+$ and $\tscrI^-$ respectively. Slices of $\tscrI^+$ and
$\tscrI^-$ are mapped by $\xi^{-1}$ to slices of $\scrI^+$ and
$\scrI^-$ respectively.

The following is a useful restriction on the causal structure of
a WASE space-time.  It is equivalent to a definition of asymptotic
simplicity in \cite{New84a} but is re-expressed here in a form
more convenient for present purposes. The change of terminology
seems appropriate because the term ``simple'' has become
overworked.

\begin{definition}\label{ACC}
A WASE asymptote $(\bM,\bg,\Omega,\psi )$ is asymptotically
chronologically consistent if $\scrU$, $(M',\g')$ and
$(\tM,\tg,\Omega',\psi')$ in Definition \ref{NWASE} may be chosen
such that for any achronal set $\tscrA$ of $(\tM,\tg )$ such that
$\tscrA \subset \psi'(\scrU )\cup \partial \tM$ one has that
$\xi^{-1}(\tscrA )\subset \psi (\scrU )\cup \partial \bM$ is
achronal in $(\bM,\bg )$.
\end{definition}

\section{The main result}

A form of weak cosmic censorship hypothesis will be required.
The H{\&}E concept of future asymptotic predictability is suitable
for this purpose.  The following definition formulates future
asymptotic predictability in terms of asymptotes and provides a
weaker concept of partial future asymptotic predictability that is
also well-established in the literature.

\begin{definition}Let $\scrS$ be a closed achronal set without edge
in a WASE space-time $(M,\g)$ and let $(\bM,\bg,\Omega,\psi)$ be a
WASE asymptote of $(M,\g)$.  Then $(\bM,\bg,\Omega,\psi)$ is future
asymptotically predictable from $\psi(\scrS)$ if one has
$\scrI^+\subset \bD^+(\psi(\scrS),\bg;\bM)$.
One says that $(\bM,\bg,\Omega,\psi)$ is partially future
asymptotically predictable from $\psi(\scrS)$ if there exists a
slice $\Sigma^+$ of $\scrI^+$ such that $J^-(\Sigma^+,\bg;\bM)\cap
\scrI^+\subset \bD^+(\psi(\scrS),\bg;\bM)$.
\end{definition}

The main result is the following:
\begin{theorem}\label{maintheorem}
Let $(\bM,\bg)$ be a WASE space-time and let
$(\bM,\bg,\Omega,\psi)$ be a WASE asymptote of $(M,\g)$.  Suppose
\begin{enumerate}[1)]
\item there exists a closed, edgeless achronal set $\scrS$ in $(M,\g)$
such that $(\bM,\bg,\Omega,\psi)$ is future asymptotically
predictable from $\psi(\scrS)$;
\item $(\bM,\bg,\Omega,\psi)$ is asymptotically chronologically
consistent;
\item one has $R_{ab}k^ak^b \geq 0$ for all null vectors $k^a$,
\end{enumerate}
then for any closed trapped surface $\scrT$ of $(M,\g)$ in
$\bI^+(\scrS,\g;M)$ one has $\psi(\scrT)\cap
J^-(\scrI^+,\bg;\bM)=\emptyset$.
\end{theorem}

Conditions (1) and (3) of Theorem \ref{maintheorem} coincide with
conditions in the statement of H{\&}E Proposition 9.2.1. However
condition (2), which would seem to be necessary, makes no
appearance in H{\&}E. Note also that Theorem \ref{maintheorem}
requires only $\scrT \subset \bI^+(\scrS,\g;M)$ whereas H{\&}E
impose the stronger condition $\scrT \subset D^+(\scrS,\g;M)$.

The basic idea of the H{\&}E argument in support of the statement
of their Proposition 9.2.1 is to show that if $\scrT$ is visible
from $\scrI^+$ then there must be a null geodesic generator of
$\dI^+(\scrT,\g;M)$ which reaches from $\scrT$ to $\scrI^+$ and
which is therefore of infinite affine length. A contradiction
then follows by means of the Raychaudhuri equation and the null
convergence condition.  The argument fails though because the
H{\&}E definition of a WASE space-time is not sufficiently
strong.  Specifically, things begin to go wrong when they claim
that, in the associated ASE space-time $(M',\g')$, for a Cauchy
surface $\scrS'$ chosen such that $\scrS'\cap \scrU'=\scrS \cap
\scrU$ it is necessarily the case that $\scrS'\setminus \scrU'$
is compact. (At this point H{\&}E are tacitly identifying $\scrU$
and $\scrU'$, as is explicitly done in the present formalism.) In
the first place it is unclear that there need be any Cauchy
surface $\scrS'$ of $(M',\g')$ such that $\scrS'\cap \scrU'=\scrS
\cap \scrU$. For example $\scrS \cap \scrU$ might not be achronal
in $(M',\g')$. And second, since $\scrS$ need not even intersect
$\scrU$ and every Cauchy surface of $(M',\g')$ is non-compact,
the set $\scrS'\cap \scrU'$ could be non-compact. Definition
\ref{NWASE} directly overcomes the second difficulty, as was
indicated in the Remark that followed Definition \ref{NWASE}. In
order to overcome the first it will be necessary to employ
different techniques.

\begin{figure}[h]
\begin{center}
\includegraphics[width=3.5in]{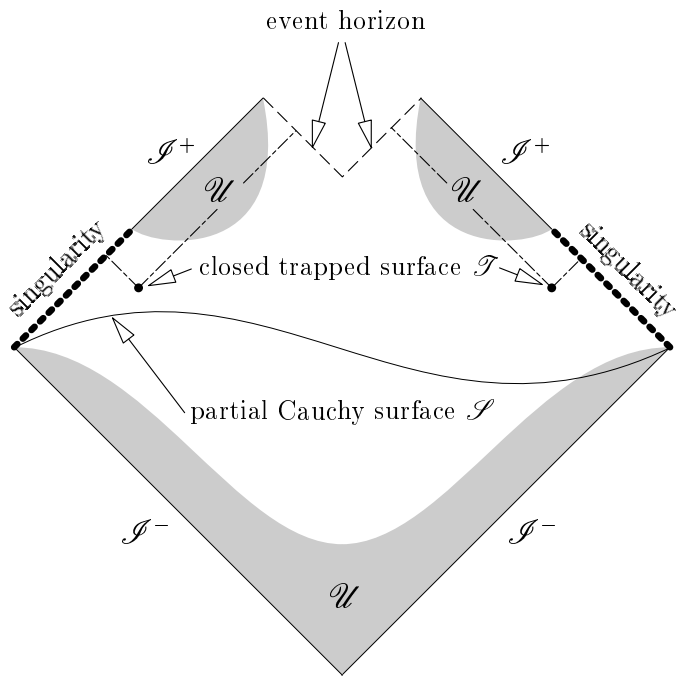}
\end{center}
\caption{A space-time which is WASE in the sense of H{\&}E and
future asymptotically predictable from a partial Cauchy surface $\scrS$.
The closed trapped surface $\scrT$ is visible from $\scrI^+$ but no null
geodesic generator of the boundary of the causal future of $\scrT$
meets $\scrI^+$.}
\end{figure}

At a pictorial level one might seek a counterexample to H{\&}E
Proposition 9.2.1 by arranging to have $\scrI^+$ both separated
from $\scrI^-$ and contained in the chronological future of a
closed trapped surface $\scrT$ (see Figure 4).  Since none of the
null geodesic generators of $\dJ^+(\psi(\scrT),\bg;\bM)$ then
meet $\scrI^+$, the central contradiction in the proof of H{\&}E
Lemma 9.2.1 is avoided. However to obtain a full counterexample
to H{\&}E Proposition 9.2.1 one also needs to arrange for the
null convergence condition to be satisfied.  Even though it is
not clear how this might be done, it seems unlikely that H{\&}E
Proposition 9.2.1 is correct.

The following three lemmas and a proposition are the key to the
proof of the revised trapped surfaces theorem.

\begin{lemma}\label{mainlemma}
Let $(M,\g )$ be a WASE space-time with a asymptotically
chronologically consistent WASE asymptote $(\bM,\bg,\Omega,\psi)$
and let $\Sigma^-$ be a slice of $\scrI^-$.  Then there exists a
slice $\Sigma^+$ of $\scrI^+$ such that $J^-(\Sigma^+,\bg;\bM)\cap
J^+(\Sigma^-,\bg;\bM)$ is a closed set of $\bM$ contained in
$\psi(\scrU)\cup \partial \bM$ and $J^-(\tSigma^+,\tg;\tM)\cap
J^+(\tSigma^-,\tg;\tM)$ is a closed set of $\tM$ contained in
$\psi'(\scrU)\cup \partial \tM$ for $\tSigma^-:=\xi(\Sigma^-)$ and
$\tSigma^+:=\xi(\Sigma^+)$.
\end{lemma}

\begin{proof}
The time reverse of Lemma 3.6 of \cite{New89b} gives that $\bM
\setminus \partial \bM$ cannot be contained entirely in
$I^+(\tSigma^-,\tg;\tM )$. Let $\tildep_+\in \tM \setminus
(I^+(\tSigma^-,\tg;\tM ) \cup \partial \tM)$ and let $\tildep \in
I^-(\tildep_+,\tg;\tM)\setminus \partial \tM$. One then has
$\tildep \in \tM \setminus (J^+(\tSigma^-,\tg;\tM ) \cup \partial
\tM)$. Let $p':= (\psi')^{-1}(\tildep)$. The set $\scrQ':=\{ p'
\}\cup(M'\setminus (\scrU \cup I(p,\g';M')))$ is compact in $M'$
by Definition \ref{NWASE} and $\tscrQ:=\psi'(\scrQ' )$ is compact
in $\tM \setminus \partial \tM$. So by Lemma 4.5 (III) of
\cite{New89b} the set $\tSigma^+_0 := \dJ^+(\tscrQ,\tg;\tM )\cap
\tscrI^+$ is a slice of $\tscrI^+$. The set
$J^-(\tSigma^+_0,\tg;\tM )\setminus (\psi'(\scrU )\cup \partial
\tM )$ is contained in $I^-(\tildep,\tg ;\tM)$ which does not
intersect $J^+(\tSigma^-,\tg;\tM )$. Hence
$J^-(\tSigma^+_0,\tg;\tM )\cap J^+(\tSigma^-,\tg;\tM )$ is
contained in $\psi'(\scrU )\cup \partial \tM$.

The set $\Sigma^+_0 := \xi^{-1}(\tSigma^+_0)$ is a slice of
$\scrI^+$. Suppose there exists a point $x \in
(I^-(\Sigma^+_0,\bg;\bM)\cap J^+(\Sigma^-,\bg;\bM))\setminus (\psi
(\scrU)\cup \partial \bM)$.  Then there exists a causal curve
$\alpha$ of $(\bM,\bg)$ from $\Sigma^-$ to $x$, and a timelike
curve $\beta :[0,1]\rightarrow \bM$ of $(\bM,\bg )$ from $x$ to
$\Sigma^+_0$. Let $a:=\sup \{ t \in [0,1]: \beta (t)\notin \psi
(\scrU)\cup \partial \bM \}$. Then $\beta (a)$ lies in the
topological boundary of $\psi (\scrU)\cup \partial \bM$ in $\bM$
and one has $\beta (t)\in (\psi (\scrU)\cup \partial \bM)\cap
I^+(\Sigma^-,\bg;\bM)\cap J^-(\Sigma^+_0,\bg;\bM)$ for all $t \in
(a,1]$. For each $t \in (a,1]$ the set $\Sigma^-\cup \{ \beta (t)
\}$ is non-achronal in $(\bM,\bg)$ and contained in
$\psi(\scrU)\cup \partial \bM$.  So, by the asymptotic
chronological consistency of $(\bM,\bg,\Omega,\psi)$, one has that
$\xi(\Sigma^-\cup \{ \beta (t) \})= \tSigma^-\cup \{\xi \circ
\beta (t)\}$ is non-achronal in $(\tM,\tg)$ for each $t \in
(a,1]$. Since $\xi \circ \beta :[a,1]\rightarrow \tM$ is a
timelike curve of $(\tM ,\tg)$ from $\xi \circ \beta (a)$ to
$\tSigma^+_0$ one thus has $\xi \circ \beta (t)\in
I^+(\tSigma^-,\tg;\tM)\cap J^-(\tSigma^+_0,\tg;\tM)$ for all $t
\in (a,1]$. Lemma 4.12 (I) of \cite{New89b} gives that
$J^+(\tSigma^-,\tg;\tM)\supset I^+(\tSigma^-,\tg;\tM)$ is closed
in $\tM$. Hence one has $\xi \circ \beta (t)\in
J^+(\tSigma^-,\tg;\tM)\cap J^-(\Sigma^+_0,\tg;\tM)$ for all $t \in
[a,1]$. But since $\beta (a)$ lies in the topological boundary of
$\psi (\scrU ) \cup \partial \bM$ in $\bM$  the point $\xi \circ
\beta (a)$ must lie in the topological boundary of $\psi'(\scrU
)\cup \partial \tM$ in $\tM$. This is impossible because
$J^-(\tSigma^+_0,\tg;\tM )\cap J^+(\tSigma^-,\tg;\tM )$ is
contained in $\psi'(\scrU )\cup \partial \tM$ which is open in
$\tM$.  One thus has $I^-(\Sigma^+_0,\bg;\bM)\cap
J^+(\Sigma^-,\bg;\bM)\subset \psi (\scrU)\cup \partial \bM$.

Now let $\tSigma^+$ be a slice of $\tscrI^+$ lying strictly to the
past of $\tSigma^+_0$ along the generators of $\tscrI^+$. One has
$J^-(\tSigma^+,\tg;\tM )\cap J^+(\tSigma^-,\tg;\tM )\subset
J^-(\tSigma^+_0,\tg;\tM )\cap J^+(\tSigma^-,\tg;\tM )\subset
\psi'(\scrU)\cup \partial \tM$.  Hence a point of $\tM$ lies in
$J^-(\tSigma^+,\tg;\tM )\cap J^+(\tSigma^-,\tg;\tM )$ iff it lies
on a causal curve of $(\tM,\tg)$ in $\psi'(\scrU)\cup \partial
\tM$ from $\tSigma^-$ to $\tSigma^+$.  Moreover, since Lemma 4.12
(I) of \cite{New89b} gives that $J^-(\tSigma^+_0,\tg;\tM )$ and
$J^+(\tSigma^-,\tg;\tM )$ are both closed in $\tM$, the set
$J^-(\tSigma^+,\tg;\tM )\cap J^+(\tSigma^-,\tg;\tM )$ is closed
in $\tM$.

Let $\Sigma^+:= \xi^{-1}(\tSigma^+)$.  One has
$J^-(\Sigma^+,\bg;\bM ) \cap J^+(\Sigma^-,\bg;\bM )\subset
(I^-(\tSigma^+_0,\bg;\bM )\cap J^+(\Sigma^-,\bg;\bM ))\cup
\partial \bM \subset \psi(\scrU)\cup \partial \bM$.  Hence a
point of $\bM$ lies in $J^-(\Sigma^+,\bg;\bM )\cap
J^+(\Sigma^-,\bg;\bM )$ iff it lies on a causal curve of
$(\bM,\bg)$ in $\psi(\scrU)\cup \partial \bM$ from $\Sigma^-$ to
$\Sigma^+$.  One thus has $J^-(\Sigma^+,\bg;\bM )\cap
J^+(\Sigma^-,\bg;\bM )= \xi^{-1}(J^-(\tSigma^+,\tg;\tM )\cap
J^+(\Sigma^-,\tg;\tM ))$. Since $J^-(\tSigma^+,\tg;\tM )\cap
J^+(\tSigma^-,\tg;\tM )$ has been shown to be closed in $\tM$ it
follows that $J^-(\Sigma^+,\bg;\bM )\cap J^+(\Sigma^-,\bg;\bM )$
is closed in $\bM$.
\end{proof}

\begin{corollary*}
Let $\scrK$ be a compact set of $M$.  Then there exists a slice
$\Sigma^+_1$ of $\scrI^+$ such that $J^-(\Sigma^+_1,\bg;\bM)\cap
J^+(\Sigma^-,\bg;\bM)$ does not intersect $\psi(\scrK)$.
\end{corollary*}

\begin{proof}
One may assume $\scrU \subset M \setminus \scrK$ otherwise one may
redefine $\scrU$ as $\scrU \setminus \scrK$.
The Lemma then gives that there exists a slice $\Sigma^+_1$ of
$\scrI^+$ such that
$J^-(\Sigma^+_1,\bg;\bM)\cap J^+(\Sigma^-,\bg;\bM)\subset
\psi(\scrU)\cup \partial \bM \subset \bM \setminus \psi(\scrK)$.
\end{proof}

\begin{lemma}\label{secondlemma}
Let $\Sigma^-$ be a slice of $\scrI^-$ and let $\scrN$ be an open neighbourhood
of $\Sigma^-$ in $\bM$.  Then there exists a slice $\Sigma^+_2$ of $\scrI^+$
such that, for every $q \in \Sigma^+_2$, the set $\scrN$ is cut by every
past endless timelike curve of $(\bM,\bg )$ to $q$.
\end{lemma}

\begin{proof}
It suffices to assume $\scrN \subset \psi(\scrU)\cup \partial \bM$
since one may otherwise redefine $\scrN$ as $\scrN
\cap(\psi(\scrU)\cup \partial \bM)$.

The set $\tscrN:=\xi(\scrN)$ is open in $\tM$ and $\tSigma^-:=
\xi(\Sigma^-)$ is a slice of $\tscrI^-$. By the time reverse of
Lemma 4.12 of \cite{New89b} one has that
$\dJ^+(\tSigma^-,\tg;\tM)$ is compact in $(\tM,\tg)$ and such that
$\dJ^+(\tSigma^-,\tg;\tM)\cap \tscrI^-=\tSigma^-$. The time
reverse of Proposition 7.2 of \cite{New89b} gives $\tscrI^+\subset
I^+(\tSigma^-,\tg;\tM)$ whereby one has
$\dJ^+(\tSigma^-,\tg;\tM)\cap \tscrI^+=\emptyset$. Hence one has
$\dJ^+(\tSigma^-,\tg;\tM)\cap \partial \tM =\tSigma^-$. Let
$\tscrX$ be the compact set $\dJ^+(\tSigma^-,\tg;\tM)\setminus
\tscrN \subset \tM \setminus \partial \tM$ or, if this is empty,
let $\tscrX$ be any non-empty compact set of $\tM \setminus
\partial \tM$. The set $\tSigma^+_{\mbox{\scriptsize$\tscrX$}}:=
\dJ^+(\tscrX,\tg;\tM)\cap \tscrI^+$ is a slice of $\tscrI^+$.
Since $I^-(\tSigma^+_{\mbox{\scriptsize$\tscrX$}},\tg;\tM)$ does
not intersect $\dJ^+(\tSigma^-,\tg;\tM)\setminus \tscrN$ one has
$\dJ^+(\tSigma^-,\tg;\tM)\cap
I^-(\tSigma^+_{\mbox{\scriptsize$\tscrX$}},\tg;\tM)\subset
\tscrN$.

Let $\Sigma^+$ be as in the statement of Lemma \ref{mainlemma} and
let $\tSigma^+:=\xi(\Sigma^+)$. Let $\tSigma^+_2$ be a slice of
$\tscrI^+$ lying strictly to the past of both
$\tSigma^+_{\mbox{\scriptsize$\tscrX$}}$ and $\tSigma^+$ along
the generators of $\tscrI^+$. One then has
$J^-(\tSigma^+_2,\tg;\tM)\cap J^+(\tSigma^-,\tg;\tM)\subset
\psi'(\scrU)\cup \partial \bM$ and $\dJ^+(\tSigma^-,\tg;\tM)\cap
J^-(\tSigma^+_2,\tg;\tM)\subset \dJ^+(\tSigma^-,\tg;\tM)\cap
(I^-(\tSigma^+_{\mbox{\scriptsize$\tscrX$}},\tg;\tM)\cup \tscrI^+)
\subset \tscrN$.

The set $\Sigma^+_2:=\xi^{-1}(\tSigma^+_2)$ is a slice of
$\scrI^+$. Let $q \in \Sigma^+_2$ and let $\sigma : (-\infty
,0]\rightarrow \bM \setminus \partial \bM$ be a future-directed,
past endless timelike curve of $(\bM,\bg)$ having a future
endpoint at $\sigma (0)=q \in \Sigma^+_2$. Let $\nu :
(b,0]\rightarrow \bM$ be the maximal segment of $\sigma$ to $q \in
\Sigma^+_2\subset \psi(\scrU)\cup \partial \bM$ in
$\psi(\scrU)\cup \partial \bM$. Then $\tnu := \xi \circ \nu$ is a
timelike curve of $(\tM,\tg)$ to $\tildeq :=\xi (q)\in
\tSigma^+_2$. One clearly has $|\tnu|\subset
J^-(\tSigma^+_2,\tg;\tM)$.  In order to show that $\tnu$ cuts
$\tscrN$ it therefore suffices to show that $\tnu$ cuts
$\dJ^+(\tSigma^-,\tg;\tM)$.

Suppose first that $\tnu$ is past endless in $(\tM,\tg)$. The time
reverse of Lemma 3.6 of \cite{New89b} gives that
$I^+(\tSigma^-,\tg;\tM)$ cannot contain all of $\bM \setminus
\partial \tM$, whilst the time reverse of Lemma 4.2 of \cite{New89b}
gives $\bM \setminus \partial \bM \subset I^+(|\nu |,\tg;\tM)$.
Hence $\tnu$ cuts $\tM \setminus I^+(\tSigma^-,\tg;\tM)$. Since
$\tnu$ is past endless and timelike in $(\tM,\tg)$ it follows
that $\tnu$ cuts $\tM \setminus J^+(\tSigma^-,\tg;\tM)$.  Because
$\tnu$ has a future endpoint at $q \in \tSigma^+\subset
\tscrI^+\subset I^+(\tSigma^-,\tg;\tM)$ one thus has that $\tnu$
cuts both $I^+(\tSigma^-,\tg;\tM) \subset J^+(\tSigma^-,\tg;\tM)$
and $\tM \setminus J^+(\tSigma^-,\tg;\tM)$ and so cuts
$\dJ^+(\tSigma^-,\tg;\tM)$.

Now suppose that $\tnu$ has a past endpoint $\tildez$ in
$(\tM,\tg)$.  The point $\tildez$ must lie in the topological
boundary of the open set $\psi'(\scrU)\cup \partial \tM$ in $\tM$
otherwise it would lie in $\psi'(\scrU)\cup \partial \tM$, in
which case $\xi^{-1}(\tildez)\in \psi(\scrU)\cup \partial \bM$
would be a past endpoint to $\nu$ in $(M,\g)$ and $\nu$ would
be past extendible in $\psi(\scrU)\cup \partial \bM$. Because
$\psi'(\scrU)\cup \partial \tM$ is open in $\tM$ the set
$J^-(\tSigma^+,\tg;\tM)\cap J^+(\tSigma^-,\tg;\tM)\subset
\psi'(\scrU)\cup \partial \tM$ cannot contain $\tildez$. Since
$J^-(\tSigma^+,\tg;\tM)\cap J^+(\tSigma^-,\tg;\tM)$ is closed in
$\tM$ the set $\tM \setminus (J^-(\tSigma^+,\tg;\tM)\cap
J^+(\tSigma^-,\tg;\tM))$ is an open neighbourhood of $\tildez$ in
$\tM$ and so is cut by $\tnu$. In view of $|\tnu|\subset
J^-(\tSigma^+_2,\tg;\tM)\subset J^-(\tSigma^+,\tg;\tM)$ one thus
has that $\tnu$ cuts $\tM \setminus J^+(\tSigma^-,\tg;\tM)$. Since
$\tnu$ has a future endpoint at $\tildeq \in \tSigma^+_2\subset
\tscrI^+\subset I^+(\tSigma^-,\tg;\tM)$ it follows that $\tnu$
cuts both $I^+(\tSigma^-,\tg;\tM)\subset J^+(\tSigma^-,\tg;\tM)$
and $\tM \setminus J^+(\tSigma^-,\tg;\tM)$ and so cuts
$\dJ^+(\tSigma^-,\tg;\tM)$.

Since $\tnu := \xi \circ \nu$ cuts $\dJ^+(\tSigma^-,\tg;\tM)$ and
therefore cuts $\tscrN:=\xi(\scrN)$ one has that $\nu$ cuts
$\scrN$. Hence $\sigma$ cuts $\scrN$.
\end{proof}

The final lemma will require the use of the following result.

\begin{proposition}\label{closedprop}
Let $(\bM,\bg,\Omega,\psi)$ be a WASE asymptote of a WASE
space-time $(M,\g)$ and let $\scrS$ be a closed edgeless achronal
set of $(M,\g)$. If $(\bM,\bg,\Omega,\psi)$ is partially future
asymptotically predictable from $\psi(\scrS)$ then $\psi(\scrS)$
is closed edgeless and achronal in $(\bM,\bg)$.
\end{proposition}

\begin{proof}
Let $(M',\g')$ be the associated ASE space-time and
$(\tM,\tg,\Omega',\psi')$ an ASE asymptote of $(M',\g')$.

If $\blambda$ was a timelike curve of $(\bM,\bg)$ from
$\psi(\scrS)\subset \bM \setminus \partial \bM$ to
$\psi(\scrS)\subset \bM \setminus \partial \bM$ then,
because $\partial \bM$ is a null hypersurface of $(\bM,\bg)$,
one would have $|\blambda|\subset \bM \setminus \partial \bM$
so there would exist a timelike curve $\lambda$ of $(M,\g)$
such that $\blambda=\psi \circ \lambda$.  But then $\lambda$
would be a timelike curve of $(M,\g)$ from $\scrS$ to $\scrS$.
This would contradict the achronality of $\scrS$ in $(M,\g)$.
Hence $\psi(\scrS)$ is achronal in $(\bM,\bg)$.

Suppose there exists $s \in \overline{\psi(\scrS)}\cap \scrI^+$.
Since $(\bM,\bg,\Omega,\psi)$ is partially future asymptotically
predictable from $\psi(\scrS)$ there exists a slice $\Sigma^+$ of
$\scrI^+$ such that $J^-(\Sigma^+,\bg;\bM)\cap \scrI^+\subset
\bD^+(\psi(\scrU),\bg;\bM)$.  In the case $s \in
J^-(\Sigma^+,\bg;\bM)\cap \scrI^+$, every past endless timelike
curve of $(\bM,\bg)$ to $s$ would cut $\psi(\scrS)$, and so
$I^+(\psi(\scrS),\bg;\bM)$ would be an open neighbourhood of $s
\in \overline{\psi(\scrS)}$ and so would intersect $\psi(\scrS)$.
This is impossible since $\psi(\scrS)$ is achronal in $(\bM,\bg)$.
So suppose $s \in J^+(\psi(\scrU),\bg;\bM)\setminus \Sigma^+$.
There exists $r \in \Sigma^+$ lying strictly to the past of $s$ on
the null geodesic generator of $\scrI^+$ through $s$. In view of
$r \in \bD^+(\psi(\scrU),\bg;\bM)$, every past endless timelike
curve of $(\bM,\bg)$ to $r$ must cut $\psi(\scrS)$.  But then
$I^+(\psi(\scrS),\bg;\bM)$ is an open neighbourhood in $\bM$ of
$r \in J^-(s,\bg;\bM)\setminus \{ s \}$ and therefore of $s \in
\scrI^+\cap \overline{\psi(\scrS)}$ and so must intersect
$\psi(\scrS)$.  So again one has a contradiction to the
achronality of $\psi(\scrS)$. One thus has
$\overline{\psi(\scrS)}\cap \scrI^+ =\emptyset$.

Suppose there exists $p \in \overline{\psi(\scrS)}\cap \scrI^-$.
Let $\Sigma^-$ be a slice of $\scrI^-$ such that $p \in \Sigma^-$.
By Lemma \ref{mainlemma} there exists a slice $\Sigma^+$ of
$\scrI^+$ such that $J^-(\tSigma^+,\tg;\tM)\cap
J^+(\tSigma^-,\tg;\tM)\subset \psi'(\scrU)\cup \partial \tM$ for
$\tSigma^-:=\xi(\Sigma^-)$ and $\tSigma^+:=\xi(\Sigma^+)$.
Since $(\bM,\bg,\Omega,\psi)$ is partially future asymptotically
predictable from $\psi(\scrS)$ one may assume that $\Sigma^+$ is
taken sufficiently far to the past in $\scrI^+$ to give
$\Sigma^+\subset \bD^+(\psi(\scrS),\bg;\bM)$.
Let $\tildep:=\xi(p)\in \tSigma^-$. By Lemma 7.2 of \cite{New89b}
one has $\tildep \in \tscrI^-\subset I^-(\tSigma^+,\tg;\tM)$.
Hence there is a timelike curve $\talpha$ of $(\tM,\tg)$ from
$\tildep \in \tSigma^-$ to some point $\tildeq \in \tSigma^+$.
In view of $|\talpha |\subset J^-(\tSigma^+,\tg;\tM)\cap
J^+(\tSigma^-,\tg;\tM)\subset \psi'(\scrU)\cup \partial \tM$ one has
that $\alpha :=\xi^{-1}\circ \talpha$ is a timelike curve of
$(\bM,\bg)$ from $p \in \overline{\psi(\scrS)}\cap \scrI^-$ to
$q:=\xi^{-1}(\tildeq)\in \Sigma^+$.
The set $I^+(p,\bg;\bM)$ cannot intersect $\psi(\scrS)$ otherwise
$I^-(\psi(\scrS),\bg;\bM)$ would be an open neighbourhood of
$p \in \overline{\psi(\scrS)}$ and so $\psi(\scrS)$ would not be achronal
in $(\bM,\bg)$.  Since the past endless null geodesic generating segment
of $\scrI^-$ to $p$ clearly does not cut $\psi(\scrS)$ it follows that
$I^+(p,\bg;\bM)$ does not intersect $D^+(\psi(\scrS),\bg;\bM)$.
But $I^+(p,\bg;\bM)$ is a neighbourhood of $q \in \Sigma^+\subset
\bD^+(\psi(\scrS),\bg;\bM)$ in $\bM$ so one has a contradiction.

One now has $\overline{\psi(\scrS)}\cap \partial \bM =\emptyset$.
Since $\psi :M \rightarrow \bM$ is a diffeomorphism onto its image
it follows that $\psi(\scrS)$ is relatively closed in $\psi(M)=
\bM \setminus \partial \bM$.  Hence $\psi(\scrS)$ is closed in
$\bM$.

One has $\edge(\psi(\scrS),\bg;\bM)\cap \partial \bM=\emptyset$
because $\psi(\scrS)$ is closed in $\bM$ and does not intersect
$\partial \bM$. And, because $\psi:(M,\g) \rightarrow (\bM,\bg)$
is a conformal isometry onto its image, one has
$\edge(\psi(\scrS),\bg;\bM)\cap \psi(M)=
\psi(\edge(\scrS,\g;M))=\emptyset$.  Hence
$\edge(\psi(\scrS),\bg;\bM)$ is empty.
\end{proof}

\begin{lemma}\label{lemma3}
Let $(\bM,\bg)$ be a WASE space-time and let
$(\bM,\bg,\Omega,\psi)$ be a WASE asymptote of $(M,\g)$.  Suppose
\begin{enumerate}[1)]
\item there exists a closed edgeless achronal set $\scrS$ in $(M,\g)$
such that $(\bM,\bg,\Omega,\psi)$ is partially future asymptotically
predictable from $\psi(\scrS)$;
\item $(\bM,\bg,\Omega,\psi)$ is asymptotically chronologically
consistent.
\end{enumerate}
Then for any compact set $\scrK \subset \bI^+(\scrS,\g;M)$ of $M$
there exists a slice $\Sigma^+_3$ of $\scrI^+$
such that $\bI^+(\psi(\scrK),\bg;\bM)\cap \scrI^+\subset
J^+(\Sigma^+_3,\bg;\bM)$.
\end{lemma}

\begin{proof}
It suffices to assume $\scrK \cap \scrU=\emptyset$ since one may
otherwise redefine $\scrU$ as $\scrU \setminus \scrK$.

Let $\Sigma^-_0$ be a slice of $\scrI^-$ and let $\Sigma^-$ be a
slice of $\scrI^-$ lying strictly to the past of $\Sigma^-_0$
along the null geodesic generators of $\scrI^-$. Let
$\tSigma^-_0:=\xi(\Sigma^-_0)$ and $\tSigma^-:=\xi(\Sigma^-)$. By
Lemma \ref{mainlemma} there exists a slice $\Sigma^+_0$ of
$\scrI^+$ such that $J^+(\Sigma^-,\bg;\bM)\cap
J^-(\Sigma^+_0,\bg;\bM)\subset \psi(\scrU)\cup \partial \bM$ and
$J^+(\tSigma^-,\tg;\tM)\cap J^-(\tSigma^+_0,\tg;\bM)\subset
\psi'(\scrU)\cup \partial \tM$ for $\tSigma^+_0:=\xi(\Sigma^+_0)$.
Lemma 4.12 of \cite{New89b} gives $\dJ^+(\tSigma^-,\tg;\tM)\cap
\scrI^-=\tSigma^-$ which, since $\tSigma^-_0$ lies strictly to
the future of $\tSigma^-$ along the null geodesic generators of
$\tscrI^-$, implies that $J^+(\tSigma^-,\tg;\tM)$ is a
neighbourhood of $\tSigma^-_0$ in $\tM$.  Proposition 7.2 of
\cite{New89b} gives $\tSigma^-_0\subset \tscrI^-\subset
I^-(\tSigma^+_0,\tg;\tM)$ whereby one has that
$J^-(\tSigma^+_0,\tg;\tM)\supset I^-(\tSigma^+_0,\tg;\tM)$ is a
neighbourhood of $\tSigma^-_0$ in $\tM$. Thus there exists an
open neighbourhood $\tscrN \subset J^+(\tSigma^-,\tg;\tM)\cap
J^-(\tSigma^+_0,\tg;\tM)$ of $\tSigma^-_0$ in $\tM$. The set
$\scrN:=\xi^{-1}(\tscrN)\subset
\xi^{-1}(J^+(\tSigma^-,\tg;\tM)\cap J^-(\tSigma^+_0,\tg;\tM))=
J^+(\Sigma^-,\bg;\bM)\cap J^-(\Sigma^+_0,\bg;\bM)$ is an open
neighbourhood of $\Sigma^-_0$ in $\bM$.  In view of Proposition
\ref{closedprop} one may, by passing to a subset of $\scrN$ if
necessary, assume $\scrN \cap (\psi(\scrS)\cup
\scrI^+)=\emptyset$. By passing to a further subset of $\scrN$ if
necessary, one may arrange that each point of $\scrN \setminus
\scrI^-$ is a future endpoint of a timelike curve of $(\bM,\bg)$
in $\scrN$ from $\scrI^-\cap \scrN$.  Lemma \ref{secondlemma}
gives that there exists a slice $\Sigma^+_2$ of $\scrI^+$ such
that every past endless timelike curve of $(\bM, \bg)$ to
$\Sigma^+_2$ cuts $\scrN$. One may assume that $\Sigma^+_2$ lies
strictly to the past of $\Sigma^+_0$ along the null geodesic
generators of $\scrI^+$. Since $(\bM,\bg,\Omega,\psi)$ is
partially future asymptotically predictable from $\psi(\scrS)$
there exists a slice $\Sigma^+_3$ of $\scrI^+$ lying strictly to
the past of $\Sigma^+_2$ along the generators of $\scrI^+$ such
that $\Sigma^+_3\subset \bD^+(\psi(\scrS),\bg;\bM)$.

Suppose there exists a timelike curve $\alpha :[0,1]\rightarrow
\bM$ of $(\bM,\bg)$ from $\psi(\scrK)\subset \bM \setminus
(\psi(\scrU)\cup \partial \bM)$ to $\Sigma^+_3$. In view of
$J^+(\Sigma^-,\bg;\bM)\cap J^-(\Sigma^+,\bg;\bM)\subset
\psi(\scrU)\cup \partial \bM$ the set $I^-(\alpha (0),\bg;\bM)$
cannot intersect $J^+(\Sigma^-,\bg;\bM)$. So, because every past
endless timelike curve of $(\bM,\bg)$ to $\Sigma^+_3$ cuts $\scrN
\subset J^+(\Sigma^-,\bg;\bM)$, there exists $a \in (0,1)$ such
that $\alpha (a)\in \scrN \setminus \scrI^-$. By the construction
of $\scrN$ there exists $\barx \in I^-(\alpha (a),\bg;\scrN)\cap
\scrI^-$. If $\alpha|[a,1]$ did not cut $\psi(\scrS)$ one could
concatenate the past endless null geodesic generating segment of
$\scrI^-\subset \bM \setminus \psi(\scrS)$ to $\barx$, a timelike
curve in $\scrN \subset \bM \setminus \psi(\scrS)$ from $\barx$
to $\alpha (a)$, and the segment $\alpha |[a,1]$ of $\alpha$ from
$\alpha (0)$ to $\alpha (1)$ to obtain a past endless causal
curve $\beta$ of $(\bM,\bg)$ to $\alpha (1)\in \Sigma^+_3$ which
did not cut $\psi(\scrS)$. For an open neighbourhood
$\scrO_{\alpha (1)}\subset \bM \setminus \overline{\psi(\scrS)}$
of $\alpha (1)\in \bM \setminus \overline{\psi(\scrS)}$ in $\bM$
there would exist $c \in (a,1)$ such that $\alpha (c)\in
\scrO_{\alpha (1)}$. But then $I^+(\alpha (c),\bg;\scrO_{\alpha
(0)})$ would be an open neighbourhood of $\alpha (1)\in
\Sigma^+_3$ in $\bM$ not intersecting $D^+(\psi(\scrS),\bg;\bM)$,
which gives a contradiction. Thus $\alpha|[a,1]$ must cut
$\psi(\scrS)$ and indeed there must exist $b \in (a,1)$ such that
$\alpha (b)\in \psi(\scrS)$. One now has that $\alpha|[0,b]$ is a
timelike curve of $(\bM,\bg)$ from $\alpha (0)\in \psi(\scrK)$ to
$\alpha (b)\in \psi(\scrS)$. Hence $I^-(\psi(\scrS),\bg;\bM)$
intersects $\psi(\scrK)\subset \bI^+(\psi(\scrS),\bg;\bM)$ and so
intersects $I^+(\psi(\scrS),\bg;\bM)$. This contradicts the
achronality of $\psi(\scrS)$ in $(\bM,\bg)$. Hence there can be
no timelike curve of $(\bM,\bg)$ from $\psi(\scrK)$ to
$\Sigma^+_3$.  One thus has $\psi(\scrK)\cap
I^-(\Sigma^+_3,\bg;\bM)=\emptyset$.

Suppose there exists $\bary \in
(\bI^+(\psi(\scrK),\bg;\bM)\cap \scrI^+)\setminus J^+(\Sigma^+_3,\bg;\bM)$.
Since $\scrI^+\setminus J^+(\Sigma^+_3,\bg;\bM)$ is relatively open in
$\scrI^+$ one can construct an open neighbourhood
$\scrO_{\bary}$ of $\bary$ in $\bM$ such that every point of
$\scrO_{\bary}\setminus \scrI^+$ is a past endpoint of a
timelike curve of $(\bM,\bg)$ in $\scrO_{\bary}$ to
$\scrI^+\setminus J^+(\Sigma^+_3,\bg;\bM)\subset J^-(\Sigma^+_3,\bg;\bM)$.
Then $I^+(\psi(\scrK),\bg;\bM)$ intersects
$\scrO_{\bary}\setminus \scrI^+\subset I^-(\Sigma^+_3,\bg;\bM)$ and so
$I^-(\Sigma^+_3,\bg;\bM)$ intersects $\psi(\scrK)$, which is
impossible. Hence $\bI^+(\psi(\scrK),\bg;\bM)$ does not intersect
$\scrI^+\setminus J^+(\Sigma^+_3,\bg;\bM)$.
There follows $\bI^+(\psi(\scrK),\bg;\bM)\cap \scrI^+
\subset J^+(\Sigma^+_3,\bg;\bM)$.
\end{proof}

It is now possible to give the proof of Theorem
\ref{maintheorem}.\vspace{1ex}

\begin{proof}[ of Theorem \ref{maintheorem}]%
One may, by passing to a subset of $\scrU$ if necessary,
assume $\scrU \cap \scrT = \emptyset$.

Suppose, for the purpose of obtaining a contradiction, that
$\psi(\scrT)\cap J^-(\scrI^+,\bg;\bM)$ is non-empty. Then
$J^+(\psi(\scrT),\bg;\bM)\cap \scrI^+$ is non-empty and so is
$I^+(\psi(\scrT),\bg;\bM)\cap \scrI^+$. By Lemma \ref{lemma3}
there exists a slice $\Sigma^+$ of $\scrI^+$ such that
$\bI^+(\psi(\scrT),\bg;\bM)\cap \scrI^+\subset
J^+(\Sigma^+,\bg;\bM)$. Since $J^+(\Sigma^+,\bg;\bM)$ is a
non-empty proper subset of $\scrI^+$ it follows that
$\bI^+(\psi(\scrT),\bg ;\bM)\cap \scrI^+$ is a non-empty proper
subset of $\scrI^+$. Hence there exists $\barq \in
\dI^+(\psi(\scrT),\bg;\bM)\cap \scrI^+ \subset
J^+(\Sigma^+,\bg;\bM)$. There exists a null geodesic generator
$\bgamma$ of $\dI^+(\psi(\scrT),\bg;\bM)$ to $\barq$ having either
a past endpoint in $\psi(\scrT)$ or no past endpoint in $\bM$. In
the former case $\bgamma$ could not be a null geodesic generating
segment of $\scrI^+$ because it would have a past endpoint in
$\psi(\scrT)\subset \bM \setminus \partial \bM$. In the latter
case $\bgamma$ could not be a null geodesic generating segment of
$\scrI^+$ because it would then cut $\scrI^+\setminus
J^+(\Sigma^+,\bg;\bM)$. Hence one has $|\bgamma|\setminus \{\barq
\}\subset \bM \setminus
\partial \bM$.

Suppose $\bgamma$ were past endless in $(\bM,\bg)$.  Then
$\bgamma$ would be a past endless causal curve of $(\bM,\bg)$ in
$\dI^+(\psi(\scrT),\bg;\bM)$. Let $\barr \in \scrI^+$ lie
strictly to the future of $\barq$ along the null geodesic
generator of $\scrI^+$ though $\barq$. Then one could deform
$\bgamma$ to the future in $(\bM,\bg)$ so as to give a past
endless timelike curve $\bgamma_+$ of $(\bM,\bg)$ to $\barr$ in
$I^+(|\bgamma|,\bg;\bM)\subset I^+(\psi(\scrT),\bg;\bM)\subset
I^+(\psi(\scrS),\bg;\bM)$. Clearly $\bgamma_+$ could not
intersect $\psi(\scrS)$ because $\psi(\scrS)$ is achronal in
$(\bM,\bg)$. Since $\bgamma_+$ is timelike curve of $(\bM,\bg)$ to
$\barr \in \bM \setminus \overline{\psi(\scrS)}$ in $(\bM,\bg)$
it follows that there would exist a neighbourhood of $\barr$ in
$\bM$ that did not intersect $D^+(\psi(\scrS),\bg;\bM)$. This
would be contrary to the future asymptotic predictability of
$(\bM,\bg,\Omega,\psi)$ from $\psi(\scrS)$. Thus $\bgamma$ must
have a past endpoint in $(\bM,\bg)$ at $\psi(\scrT)$.
Consequently there exists a null geodesic $\gamma$ of $(M,\g)$
such that $\psi \circ \gamma$ is the unique maximal segment of
$\bgamma$ in $\bM \setminus \partial \bM$.

For each $p \in |\gamma|$, every open neighbourhood of
$\barp:=\psi(p)$ in $\bM$ intersects both
$I^+(\psi(\scrT),\bg;\bM)\setminus \partial \bM=
\psi(I^+(\scrT,\g;M))$ and $\bM \setminus
I^+(\psi(\scrT),\bg;\bM) = \psi (M \setminus I^+(\scrT,\g;M))$.
One thus has $|\gamma|\subset \dI^+(\scrT,\g;M)$ and hence that
$\gamma$ is a null geodesic generator of $\dI^+(\scrT,\g;M)$.
Since $\bgamma$ has a past endpoint at $\psi(\scrT)$ in
$(\bM,\bg)$ it follows that $\gamma$ has a past endpoint at
$\scrT$ in $(M,\g)$.  Since $\bgamma$ has a future endpoint at
$\barq \in \scrI^+$ in $(\bM,\bg)$ it follows that $\gamma$ is
future endless and future complete in $(M,\g)$.  One may assume
that $\gamma$ is an affine future-directed null geodesic of
$(M,\g)$ of the form $\gamma :[0,\infty )\rightarrow M$.

Let ${\bf k}$ and ${\bf l}$ be null normal fields to $\scrT$
along a relative open neighbourhood $\scrV_{\gamma(0)}$ of
$\gamma(0)$ in $\scrT$, normalised such that $g_{ab}k^al^b=-1$,
with ${\bf k}(\gamma(0))=\dgamma (0)\in T_{\gamma(0)}M$. The
induced metric on $\scrT$ is given by
$h_{ab}=g_{ab}+2k_{(a}l_{b)}$, whilst $\preone \chi_{ab}:= h_a
\Dummy^c h_b \Dummy^d k_{c;d}$ and $\pretwo \chi_{ab}:= h_a
\Dummy^c h_b \Dummy^d l_{c;d}$ are null second fundamental forms
of $\scrT$ along $\scrV_{\gamma(0)}\subset \scrT$. By the
definition of a closed trapped surface one has $\preone \chi^a
\dummy_a<0$ and $\pretwo \chi^a \dummy_a<0$ along
$\scrV_{\gamma(0)}$. The vector field ${\bf k}$ along
$\scrV_{\gamma(0)}\subset \scrT$ defines a congruence of future
endless affine null geodesics of $(M,\g)$ from $\scrT$ with
tangents that coincide with ${\bf k}$ along
$\scrV_{\gamma(0)}\subset \scrT$.  Let ${\bf k}$ also denote the
tangents to these null geodesics. Each tangent vector to $\scrT$
at $\gamma(0)$ may be Lie propagated along $\gamma$ with respect
to ${\bf k}$ to yield a vector field ${\bf Z}$ along $\gamma$.
From vanishing torsion one has $Z^a \dummy_{;b}k^b=k^a
\dummy_{;b}Z^b$ and hence that ${\bf Z}$ satisfies the defining
equation $(Z^a \dummy_{;b}k^b)\Dummy_{;c}k^c=R^a
\dummy_{bcd}k^bk^cZ^d$ for a Jacobi field along $\gamma$. Note
that ${\bf Z}$ is orthogonal to ${\bf k}$ at $\gamma (0)$ and
satisfies $(k_aZ^a)_{;b}k^b=0$ along $\gamma$, and so is
orthogonal to ${\bf k}$ along $\gamma$.

One may parallelly propagate the vector ${\bf l}$ along the
integral curves of ${\bf k}$ and so define
$h_{ab}=g_{ab}+2k_{(a}l_{b)}$ along these curves.  Then $h
\Dummy^a \Dummy_b$ is a projection operator such that $h \Dummy^a
\dummy_{b;c} k \Dummy^c=0$. One thus has that
$\Dummy^{\top}\hspace{-1ex}Z^a:= h \Dummy^a \dummy_bZ^b$
satisfies $\Dummy^{\top}\hspace{-1ex}Z^a \dummy_{;b}k^b= h
\Dummy^a \dummy_b k^b \dummy_{;c}\Dummy^{\top}\hspace{-1ex}Z^c =
\preone \chi \Dummy^a \dummy_b \Dummy^{\top}\hspace{-1ex}Z^b$ for
$\preone \chi_{ab}:= h_a \Dummy^ch_b \Dummy^dk_{c;d}$ now defined
all along $\gamma$. One also has $(\Dummy^{\top}\hspace{-1ex}Z^a
\dummy_{;b}k \Dummy^b)\dummy_{;c}k \Dummy^c= h \Dummy^a \dummy_b
R \Dummy^b \dummy_{cde} k \Dummy^ck
\Dummy^d\Dummy^{\top}\hspace{-1ex}Z^e$.

The expansion and shear tensors of the vector fields
$\Dummy^{\top}\hspace{-0.6ex}{\bf Z}$ along $\gamma$ may be
expressed as $\vartheta_{ab}=h_a \Dummy^ch_b \Dummy^dk_{c;d}$ and
$\varsigma_{ab}=\vartheta_{ab}-\frac{1}{2}\vartheta h_{ab}$
respectively, where $\vartheta:= h^{ab}\vartheta_{ab}$ is the
scalar expansion. Then, defining $\varsigma^2 :=
\frac{1}{2}\varsigma_{ab}\varsigma^{ab}$, one has that
$\vartheta(\lambda)$ satisfies the Raychaudhuri equation
\begin{align}
\frac{d}{d \lambda}\vartheta (\lambda) &= -R_{ab}k^ak^b
-2\varsigma^2(\lambda)-\smallfrac{1}{2}\vartheta^2(\lambda)
\label{Ray}
\intertext{and is subject to the initial condition}
\vartheta (0) &=\chi^a \dummy_a (\gamma(0)) < 0 \ .
\end{align}
By means of condition (3) of Theorem \ref{maintheorem} one thus has
\begin{equation}
\frac{d}{d \lambda} \left( \frac{2}{-\vartheta (\lambda)}\right)
\leq -1
\end{equation}
for all $\lambda \in [0,\infty)$ such that $\vartheta (\lambda)
\neq 0$.  Hence there exists
$\lambda_0\in (0,2/(-\chi^a \dummy_a(\gamma(0)))]$ such
that $\lim_{\lambda \nearrow \lambda_0}\vartheta(\lambda)=-\infty$.
Thus $\gamma (\lambda)$ is conjugate to $\scrT$ at $\lambda=\lambda_0$
and so there exists a Jacobi field ${\bf Z}$ along $\gamma$ which is
non-zero and tangent to $\scrT$ at $\gamma(0)\in \scrT$ and such that
$\Dummy^{\top}\hspace{-0.6ex}{\bf Z}(\lambda_0)=0$.
One may, by an adaptation of the technique of H{\&}E Proposition
4.5.12, use the vector field $\Dummy^{\top}\hspace{-0.6ex}{\bf Z}$
to construct a timelike curve of $(M,\g)$ from $\scrT$ to
$\gamma(\lambda)$ for any $\lambda>\lambda_0$.
But this is impossible because $\gamma$ is a
generator of $\dI^+(\scrT,\g;M)$.
This establishes the required contradiction.
\end{proof}

This then is the revised proof of the familiar assertion that,
subject to the null convergence condition and weak cosmic
censorship, closed trapped surfaces are not visible from
$\scrI^+$. The key idea of the H{\&}E argument in support of this
assertion was evidently sound, but additional constraints and
analysis have been seen to be necessary to make the detailed
theory of WASE space-times match intuitive expectations.

One could, as discussed in {\S}\ref{prelimsec}, consider weakening
the assumed strong causality of the reference ASE asymptote
$(\tM,\tg,\Omega',\psi')$ to a the chronology condition on the
underlying ASE space-time $(M',\g')$. This would lead to a more
general definition of a WASE space-time. It would be of some
interest to find whether or not Theorem \ref{maintheorem} remains
true in this more general setting.

\section{Concluding remarks}

The definition of a WASE space-time, proposed in \cite{New89b} as
the foundation for certain types of cosmic censorship theorems,
has been seen also to provide the basis for a rigorous proof of
standard relativity folklore concerning the invisibility of
closed trapped surfaces from $\scrI^+$.  Other approaches to the
definition of WASE space-time have been attempted
\cite{Kr88,WaGR84}, but a comparison will not be attempted here.

Whilst this paper was in preparation, Chru{\'s}ciel et al.\
\cite{CDGH00} have been reconsidering another piece of relativity
folklore, namely the area theorem for black holes. This states
that, subject to weak cosmic censorship and an energy condition,
the area of a black hole cannot decrease. A formalised statement
to this effect appears as Proposition 9.2.7 in H{\&}E, but the
proof there is flawed because it is based on unsubstantiated
assumptions concerning the smoothness of the event horizon. A
proof is provided in \cite{CDGH00} of a weaker area theorem, but
with full attention to matters of differentiability. A central
hypothesis is one of $\scrH$-regularity which, in present
terminology, requires that there exists a neighbourhood $\scrO$
of the event horizon $\scrH := \dJ^-(\scrI^+,\bg;\bM)$ in $\bM$
such that, for any compact set $\scrC \subset \scrO$ which
intersects $I^-(\scrI^+,\bg;\bM)$, there is a null geodesic
generator of $\scrI^+$ which cuts both $\bI^+(\scrC,\bg;\bM)$ and
$\bM \setminus \bI^+(\scrC,\bg;\bM)$.  The authors consider
various ways to derive $\scrH$-regularity from seemingly more
natural hypotheses. However, Lemma \ref{lemma3} of the present
paper shows that for space-times which are WASE in the sense of
\cite{New84a} and asymptotically chronologically consistent in
the sense of Definition \ref{ACC}, $\scrH$-regularity is in fact
a consequence of partial future asymptotic predictability.

\section*{acknowledgements}
I am grateful to Piotr Chru{\'s}ciel for having prompted me to
write this article, and for his subsequent comments. Financial
support was provided by the NFR (Natural Sciences Research
Council) of Sweden.


\bibliographystyle{plain}

\end{document}